\documentclass[aps, nofootinbib,superscriptaddress]{revtex4}
\usepackage{graphicx}
\usepackage{amsfonts}
\usepackage{psfrag}
\usepackage{amsmath}
\usepackage{float}
\usepackage{amssymb}  

\def\be{\begin{equation}}
\def\ee{\end{equation}}
\def\ba{\begin{eqnarray}}
\def\ea{\end{eqnarray}}

\def\ut#1{\rlap{\lower1ex\hbox{$\sim$}}#1{}}

\DeclareFontFamily{U}{rsfs}{}         
\DeclareFontShape{U}{rsfs}{m}{n}{<5> rsfs5 <6><7> rsfs7          %
  <8><9><10><10.95><12><14.4><17.28><20.74><24.88> rsfs10}{}     %
\DeclareMathAlphabet{\mathfs}{U}{rsfs}{m}{n}                     %
\newcommand{\mfs}[1]{\mathfs {#1}}                               %
\newcommand{\n}{{\nonumber}}
\newcommand{\sG}{{\mfs G}}

\begin{document}

\title{A note on the Poisson bracket of 2d smeared fluxes in loop quantum gravity}

\author{Alberto S. Cattaneo}
\affiliation{Institut f\"ur Mathematik, Universit\"at Z\"urich\\
Winterthurerstrasse 190, CH-8057 Z\"urich, Switzerland}
\author{Alejandro Perez}
\affiliation{
Centre de Physique Th\'eorique,   Aix Marseille Universit\'e, Universit\'e de Toulon, CNRS, UMR 7332, 13288 Marseille, France.
}
\begin{abstract}
We show that the non-Abelian nature of geometric fluxes---the corner-stone in the definition of quantum geometry in the framework of loop quantum gravity (LQG)---follows directly form the continuum
canonical commutations relations of gravity in connection variables and the validity of the Gauss law. The present treatment simplifies previous 
formulations and thus identifies
more clearly the root of the discreteness
of geometric operators in LQG. Our statement generalizes to arbitrary gauge theories and relies only on the validity of the Gauss law. 
\end{abstract}
\maketitle


The phase space of general relativity admits a parametrization in terms of an $SU(2)$ connection $A_a^i$ (called the Ashtekar-Barbero connection)
and a canonically conjugated two-form $E^i_{ab}=\epsilon^{i}_{\ jk} e^j\wedge e^k$---defined in terms of the frame one-forms $e_a^i$ encoding 
the geometry of the spacelike hypersurface $\Sigma$ that are proper to the $3+1$ Hamiltonian  decomposition of spacetime  \cite{Rovelli:2004tv, Thiemann:2007zz}. The commutation relations are
\ba
 \{E_{ab}^i(x),E_{cd}^j(y)\}&=&0 \n \label{uno} \\
\{A_{a}^i(x),A_{b}^j(y)\}&=&0\n \label{dos}  \\
 \{E_{ab}^i(y), A_{c}^j(y)\}&=&\kappa \gamma \epsilon_{abc} \delta^{ij}\delta^{(3)}(x,y), \label{tres}
\ea
where $\gamma$ is the Barbero-Immirzi parameter \cite{Barbero:1994ap, Immirzi:1996di}.
Phase space fields must satisfy first class constraints associated with the gauge symmetry content of general relativity in connection variables.
Here we only explicitly use 
the Gauss constraint\be\label{gauss}
d_A E^i=0,
\ee 
which arises from the local $SU(2) $ gauge symmetry associated with the freedom of choosing the frame field $e_a^i(x)$ on $\Sigma$. 
There is a basic set of functionals of $E^i_{ab}$ that enter the construction of loop quantum gravity. These are  given by the flux operators across a 2-dimensional surface $S\subset \Sigma$ which are defined as
\be\label{fluxes}
E_{lqg}(S,\alpha)=\int_S \alpha_iE^i,
\ee
where  $\alpha^i$ is a smearing field, here assumed to have compact support in $\Sigma$.
In apparent contradiction with \eqref{uno} the $E_{lqg}(S,\alpha)$  do not Poisson commute {\em in a suitable sense},   namely 
 \be\label{key}
\left\{E_{lqg}(S,\alpha),E_{lqg}(S,\beta)\right\}\approx\kappa \gamma E_{lqg}[S,[\alpha,\beta]],
\ee
where $[\alpha,\beta]_i=\epsilon_{ijk} \alpha^j\beta^k$ and the $\approx$ sign is used in order to emphazise the central role of the Gauss
constraint on the appearance of such non-Abelian structure.

{A key point of these notes is to point out the fact that the definition (\ref{fluxes}) is singular in the sense that these observables 
do not generate a well defined Hamiltonian vector field in phase space. However, equation (\ref{key}) can still be given a meaning by noticing that there exist a closely related well-defined 
family of observables $E(S, \alpha)$ (in the sense that they are generators on non-singular Hamiltonian vector fields in phase space) that coincide with the $E_{ lqg}(S,\alpha)$ on the constraint surface defined by \eqref{gauss}; i.e. $E_{ lqg}(S,\alpha)\approx E(S,\alpha)$.} 
Moreover,   for all well defined gauge invariant observables $O$ one has that
\be\label{equiv}
\{E_{lqg}(S,\alpha),O\}=\{E(S,\alpha),O\}.
\ee
 As the gauge non-invariant fluxes \eqref{fluxes} are of interest in LQG only because they can be used as fundamental building blocks in the construction of the gauge invariant quantum geometry operators (such as area and volume \cite{Ashtekar:1996eg, Ashtekar:1997fb}), the previous property amounts for equivalence between $E_{ lqg}(S,\alpha)$ and $E(S,\alpha)$ in physical applications. 

\subsubsection*{ Non-commutativity of fluxes}
Here we show that (\ref{key}) is a simple consequence of the fact that the fluxes are 2-dimensional smearings of a 3-dimensional distributional field.
More precisely, { is just the singular character of the smearing \eqref{fluxes} that misleads to the naive inappropriate expectation that these fluxes should 
commute}. Equations (\ref{tres}) define the Poisson brackets between suitably smeared observables in 
three dimensions. Thus, it is only after expressing (\ref{fluxes}) { in terms of well defined Hamiltonian functions---which involves an integral of a local density on a 3-dimensional region---that its genuine dependence of the 
connection becomes discernible}. Moreover, we will also uncover a clear geometric meaning of (\ref{key}) emerging naturally from our perspective. 

Thus, the key is to provide a definition of the 2-dimensional integral in (\ref{fluxes})  in terms of a 3-dimensional observable which is functionally differentiable in the phase space and coincides with the singular 2d smearing on the constraint surface defined by \eqref{gauss}. 
We assume, without loss of generality, that $S$ is a closed
 surface---if the $2$-surface $S$ were not closed we could always extend it to a new surface $S'$ in some arbitrary way in the region outside the support of $\alpha$ to have it closed in a way such that  $E_{lqg}(S,\alpha)=E_{lqg}(S',\alpha)$.  
Now we define a new quantity $E(S,\alpha)$ as
\ba\label{good}
\n E(S,\alpha)&\equiv& \int_{{\rm Int}[S]} d_A\alpha_i \wedge E^i\\
&\approx& \int_{S} \alpha_i E^i=  E_{lqg}(S,\alpha),
\ea 
where ${\rm Int}[S]\subset \Sigma$ is the region with $S$ as its boundary,  and in the second line the symbol $\approx$ reminds us that we have used the Gauss law \eqref{gauss}.
Thus, on the constraint surface,  $E(S,\alpha)$  coincides with  $E_{lqg}(S,\alpha)$: the singular two dimensional smearing  \eqref{fluxes} of the $E^i$. 
One has succeeded in writing the flux variables as the three dimensional smearing of local fields,  now one can proceed and safely compute the Poisson bracket between different fluxes. Direct calculation using (\ref{tres}) yields
\ba\label{kkey}
\left\{E(S,\alpha),E(S,\beta)\right\}&=& \int\int dx^3 dy^3 \left\{d\alpha_i\wedge E^i+\epsilon_{ijk} A^j\wedge \alpha^k\wedge E^i, 
d\beta_l\wedge E^l+\epsilon_{lmn} A^m\wedge \beta^n\wedge E^l\right\}\n \\
&=& \kappa \gamma \int  dx^3  d_A([\alpha,\beta])_k\wedge E^k\n\\
&=& \kappa \gamma E[[\alpha,\beta], S],  
\ea
where $[\alpha,\beta]_k\equiv \epsilon_{kij} \alpha^i\beta^k$. This is the sought result.

\subsubsection{Relation with the generators of gauge transformations in the presence of boundaries}

Here we make an observation that leads to a clear geometric interpretation of the observables $E(S,\alpha)$.
This observation uses a general feature of gauge systems in the presence of boundaries that was first noticed in the context of gravity some time ago \cite{Benguria:1976in}.
The observation is that in the presence of boundaries the smeared Gauss constraint
\be
\sG_R(\alpha)\equiv\frac{1}{\kappa \gamma} \int_{R\subset \Sigma} \alpha_i \, d_AE^i 
\ee
fails to be functionally differentiable due to the appearance of boundary terms.  More precisely, computing its variation we get
\ba
\kappa \gamma \delta \sG_R(\alpha)&\equiv& \int_{R\subset \Sigma} \alpha_i \, d_A (\delta E^i)+\alpha_i [\delta A,E]^i  \n \\
&=& \int_{R\subset \Sigma} -d_A \alpha_i \wedge  \delta E^i+\alpha_i [\delta A,E]^i+\int_{\partial R} \alpha_i \delta E^i,
\ea
with the appearance of a boundary contribution breaking functional differentiability.
We can relate the above generator to a functionally differentiable one $\bar \sG_R(\alpha)$ by appropriately subtracting a boundary term as follows: \ba \bar \sG_R(\alpha)&\equiv&\sG_R(\alpha)-\frac{1}{\kappa\gamma}\int_{\partial R} \alpha_i E^i\n \\ &=& -\frac{1}{\kappa\gamma}\int_{R} (d_A\alpha_i)\wedge E^i.\ea
The generators $\bar\sG_R(\alpha)$ satisfy the $SU(2)$ local gauge algebra \be \{\bar \sG_R(\alpha),\bar \sG_R(\beta)\}=- \bar \sG_R([\alpha,\beta]).\ee However, in contrast with 
the Gauss law---which must vanish on-shell of the first class constraints $\sG_R(\alpha)\approx 0$ for all suitable smearings---$\bar \sG_R(\alpha)$ does not need to vanish \footnote{The algebra (\ref{kkey}) can be completed to the holonmy-flux algebra in a standard way; see for instance Section 9.1.1 in \cite{Thiemann:2007zz} where the quantization of $\bar \sG_R(\alpha)$ is defined.}. In fact it represents a non trivial $SU(2)$ charge on the boundary which is precisely related to the flux variables \eqref{good} by the equation
\be
E(\partial R,\alpha)= -\kappa \gamma\bar \sG_R(\alpha).
\ee
The previous equation explains why the algebra (\ref{key}) is directly related to the symmetry algebra associated with the generators $\bar \sG_R(\alpha)$ of the $SU(2)$ gauge symmetry in the presence of a boundary. This completes the presentation of the statements in these notes in the context of LQG.  We hope this clarifies the geometric  origin for the non-commutativity of fluxes in loop quantum gravity which is especially important as  can be shown to be the root from which discreteness of quantum geometry stems \cite{Rovelli:1994ge,Ashtekar:1996eg, Ashtekar:1997fb}.

Of course the status of \eqref{key} is well understood from previous works; however, we believe 
our demonstration---in the form of the  off-shell algebra \eqref{kkey} and in view of \eqref{equiv}---sheds a more direct light on the simple nature of this property and its relationship with the symmetry algebra.
For instance, in reference \cite{Ashtekar:1998ak} the authors study the commutation relations of an ensemble of observables
that also includes the smearing of the connection $A_a^i$ along one dimensional paths via holonomies, and show that \eqref{key} 
follows from the consistency requirements imposed by the Jacobi identity. This is a beautiful result but it makes strong use of the commutation relations between fluxes and holonomies and thus somewhat obscures the simple nature of \eqref{key} which is valid with no need of discretization.
In a more recent work \cite{Freidel:2011ue} the authors present a modification of the definition
\eqref{fluxes} that explicitly depends on the connection $A_a^i$ in order to give a `more natural' account of \eqref{key}.
This analysis is certainly a valid way to proceed in the definition of fluxes in a gauge theory. Nevertheless, we believe that our present treatment is simpler when 
aiming at illustrating the source of non-commutativity in quantum-geometry. One additional reason for this is that, historically, 
the definition \eqref{fluxes} has been the most generally applied in LQG. 

\subsubsection*{Another example: the Poisson sigma model}

We have emphasized the role of the previous mathematical objects in the context of loop quantum gravity.
The reason is that the definition (\ref{fluxes}) is crucial in the construction of the kinematical structures that
are at the foundations of the Hilbert space of the theory that carry a unitary representation of spatial diffeomorphisms \cite{Lewandowski:2005jk}. However, our remark is general: it applies to lattice gauge theories and lies at the heart of the discreteness of the electric 
field spectrum for gauge theories with a compact gauge group. Indeed, the above construction is pretty general. Other interesting, nontrivial examples occur in $BF$ theories \cite{Schwarz:1978cn, Blau:1989bq}
and in the Poisson sigma model (PSM) \cite{Ikeda:1993fh, Schaller:1994es}.  As the case of $BF$ theory is very similar to what we have discussed above, we would like to focus on the PSM to have an example of different nature. 

The PSM is a sigma model with target a Poisson manifold. To simplify the description, we work in coordinates (that is, we assume the target to be $\mathbb{R}^n$).
The target data is
a bivector field $\pi$, i.e., a collection of skew symmetric matrices $\pi^{ij}(x)$ depending smoothly on $x\in \mathbb{R}^n$ and satisfying
\begin{equation}\label{Poisson}
\pi^{ij}\partial_i\pi^{kl}+\pi^{il}\partial_i\pi^{jk}+\pi^{ik}\partial_i\pi^{lj}
=0.
\end{equation}
The fields of the PSM on a surface $\Sigma$, with local coordinates $(\sigma^1,\sigma^2)$, are a set of functions $X^i$ and of one-forms $\eta_{i}=\eta_{i\mu}d\sigma^\mu$, $i=1,\dots,n$. The action is
\[
S=\int_\Sigma \left(\eta_{i\mu}\partial_\nu X^i +\frac12\pi^{ij}(X)\eta_{i\mu}\eta_{j\nu}\right)
d\sigma^\mu\wedge d\sigma^\nu.
\]
A boundary component of a surface is a circle. It is however more interesting to fix boundary conditions $\eta\equiv0$ on part of the boundary, so the remaining boundary components are intervals.
 Let us fix an interval, say $[0,1]$. If we denote by $\zeta_i(t)dt$ the restriction to the boundary of $\eta_i$,
 the boundary fields are functions $X^i$ and $\zeta_i$ on $[0,1]$ with
canonical commutation relations $\{\zeta_i(t_1),X^j(t_2)\}=\delta_i^j\delta(t_1-t_2)$ (all other brackets vanishing)
and Gauss constraints
\begin{equation}\label{GPSM}
\dot X^i + \pi^{ij}(X)\zeta_j=0.
\end{equation}
An interesting observable is $Y_\text{PSM}(\alpha):=\alpha_iX^i(0)$, where $\alpha$ is a vector in $\mathbb{R}^n$.
We might naively expect $\{Y_\text{PSM}(\alpha),Y_\text{PSM}(\beta\})$ to vanish but this is incorrect \cite{Cattaneo:2000iw}. In fact, using the Gauss constraints we may write
\[
Y_\text{PSM}(\alpha)=-\int_0^1d(\alpha_iX^i)
\approx
\int_0^1(\alpha_i\pi^{ij}(X)\zeta_j-\dot\alpha_iX^i)dt =:Y(\alpha),
\]
where now $\alpha$ denotes a function on $[0,1]$ that reduces to the original constant $\alpha$ at $0$ and vanishes and $1$. We now have
\ba
\n
\{Y(\alpha),Y(\beta)\} &=&
\iint \{ (\alpha_i\pi^{ij}(X)\zeta_j-\dot\alpha_iX^i)(s), (\beta_k\pi^{kl}(X)\zeta_l-\dot\beta_kX^k)(t)\}
ds\,dt=\\
&=&
\int_0^1[
\dot\alpha_i\beta_k\pi^{ki}(X)
-\alpha_i\pi^{ij}(X)\dot\beta_j
+\alpha_i\beta_k(
\pi^{ij}(X)\partial_j\pi^{kl}(X)\zeta_l - \zeta_j\partial_l\pi^{ij}(X)\pi^{kl}(X)
)
]dt.
\ea
By \eqref{Poisson} and \eqref{GPSM}, and also renaming the indices, we then have
\ba
\n
\{Y(\alpha),Y(\beta)\} &=&
-\int_0^1
(\dot\alpha_i\beta_k\pi^{ik}(X)+
\alpha_i\dot\beta_k\pi^{ik}(X)+
\alpha_i\beta_k\pi^{lj}(X)\partial_j\pi^{ik}(X)\zeta_l)dt\\
&\approx&
-\int_0^1(
\dot\alpha_i\beta_k\pi^{ik}(X)+
\alpha_i\dot\beta_k\pi^{ik}(X)+
\alpha_i\beta_k\dot X^j\partial_j\pi^{ik}(X)
)dt\\
&=&
-\int_0^1 d(\alpha_i\beta_k\pi^{ik}(X)) = 
\alpha_i(0)\beta_k(0)\pi^{ik}(X(0)),
\ea
which in general does not vanish. (Roughly speaking we may read this formula by saying that the correct
value of the bracket $\{X^i(0),X^k(0)\}$ is $\pi^{ik}(X(0))$.)

Note that in case the Poisson bivector field is linear---i.e., $\pi^{ij}(x)=f^{ij}_kx^k$ with $f^{ij}_k$ the structure constants of a Lie algebra---we are back to the case of $BF$ theory and we can write the above Poisson bracket in the form
$\{Y(\alpha),Y(\beta)\} = Y([\alpha,\beta])$, where $[\alpha,\beta]$ denotes the Lie bracket of the Lie algebra elements
$\alpha$ and $\beta$ (i.e., $[\alpha,\beta]_i = f^{jk}_i\alpha_j\beta_k$).

We conclude with a brief general explanation of the above phenomena. Let us first recall the usual framework of symplectic geometry. Here one starts with a closed, nondegenerate two-form $\omega$ (e.g., the usual $d p_i dq^i$) and associates to every function $f$ its Hamiltonian vector field $X_f$ via the equation
\begin{equation}\label{Ham}
\iota_{X_f}\omega = d f.
\end{equation}
Nondegenerate means precisely that for every $f$ there is a unique $X_f$ solving this equation.
(In the usual case $\omega=d p_i dq^i$ we get $X_{f,i}dq^i-X_f^idp_i=\partial_ifdq^i+\partial^ifdq_i$ and hence
$X_{f,i}=\partial_if$, $X_f^i=-\partial^if$.)
The Poisson bracket $\{f,g\}$ of two functions $f$ and $g$ is then defined as $X_f(g)$, or equivalently as
$-X_g(f)$ or as $\iota_{X_f}\iota_{X_g}\omega$. 

We now move to the case when $\omega$ is possibly degenerate \cite{Gotay:1978dv}. In this case, a function $f$ such that
\eqref{Ham} has solution is called Hamiltonian and a vector field $X_f$ solving it is called a Hamiltonian vector field for $f$. In the infinite dimensional case there is also the interesting possibility of a weakly nondegenerate form (which is what
usually occurs on spaces of fields) with the property that the Hamiltonian vector field is uniquely determined, when it exists. If $f$ and $g$ are Hamiltonian we still have $\{f,g\}:=X_f(g)=\iota_{X_f}\iota_{X_g}\omega=-X_g(f)$,
no matter which Hamiltonian vector fields we choose. Also note that $\{f,g\}$ is also Hamiltonian
(with Hamiltonian vector field, e.g., $[X_f,X_g]$).

Let us now suppose that we have first class constraints, i.e., Hamiltonian functions $\phi_\mu$ satisfying
$\{\phi_\mu,\phi_\nu\}=f_{\mu\nu}^\rho\phi_\rho$ for some structure functions $f_{\mu\nu}^\rho$. 
We denote by $X_{\phi_\mu}$ (a choice of) the Hamiltonian vector fields of the functions $\phi_\mu$ 
(recall however that in the weakly nondegenerate case this choice is unique).
A function $f$
is called invariant if $X_{\phi_\mu}f=f_{\mu}^{\nu}\phi_\nu$ for some functions $f_{\mu}^{\nu}$. Note that if $f$ is invariant, then so is $\Tilde f = f + f^\mu\phi_\mu$ for every choice of functions $f^\mu$. We consider $f$ and $\Tilde f$ as equivalent. Note that, if $f$ and $g$ are invariant and Hamiltonian, then their Poisson bracket $\{f,g\}$ is also invariant; moreover, changing $f$ and $g$ to equivalent Hamiltonian functions changes $\{f,g\}$ to an equivalent function. Now the important point is that equivalence does not always preserve the property of being Hamiltonian. In particular, it is possible that a non Hamiltonian, invariant function $f$ is equivalent to a Hamiltonian (and automatically invariant) function $\Tilde f$. We may then replace $f$ with $\Tilde f$ if we want to compute Poisson brackets, and the particular choice of $\Tilde f$, as long as it is Hamiltonian, is irrelevant.

In the examples coming from field theory, the form $\omega$ is local (and weakly nondegenerate). This in particular means that local functionals are Hamiltonian. However, nonlocal functionals such as those obtained by integrating on a submanifold (as in the above examples) are not Hamiltonian (their Hamiltonian vector field may formally be defined
in terms of a delta function with support on the submanifold, but this is not a smooth vector field on the space of fields; moreover, applying a distributional vector field such as this one to a non local functional may generally lead to a product of delta functions). The general trick, as explained above, is to replace the nonlocal functional at hand (if possible) by
an equivalent local one.

\section*{Acknowledgments} 

We would like to thank the organisers of the conference {\em Estate Quantistica 2014} where the present project was initiated. This research was supported  (partly) by the NCCR SwissMAP, funded by the Swiss National Science Foundation, and by the 
COST Action MP1405 QSPACE, supported by COST (European Cooperation in Science and Technology) as well as  by the OCEVU Labex (ANR-11-LABX-0060),  and (partly) by  the A*MIDEX project (ANR-11-IDEX-0001-02) funded by the ``Investissements d'Avenir" French government program managed by the ANR.

%

\begin{thebibliography}{10}

\bibitem{Rovelli:2004tv}
Carlo Rovelli.
\newblock {Quantum gravity}.
\newblock 2004.

\bibitem{Thiemann:2007zz}
Thomas Thiemann.
\newblock {Modern canonical quantum general relativity}.
\newblock 2001.

\bibitem{Barbero:1994ap}
J.~Fernando Barbero~G.
\newblock {Real Ashtekar variables for Lorentzian signature space times}.
\newblock {\em Phys. Rev.}, D51:5507--5510, 1995.

\bibitem{Immirzi:1996di}
Giorgio Immirzi.
\newblock {Real and complex connections for canonical gravity}.
\newblock {\em Class. Quant. Grav.}, 14:L177--L181, 1997.

\bibitem{Ashtekar:1996eg}
Abhay Ashtekar and Jerzy Lewandowski.
\newblock {Quantum theory of geometry. 1: Area operators}.
\newblock {\em Class.Quant.Grav.}, 14:A55--A82, 1997.

\bibitem{Ashtekar:1997fb}
Abhay Ashtekar and Jerzy Lewandowski.
\newblock {Quantum theory of geometry. 2. Volume operators}.
\newblock {\em Adv.Theor.Math.Phys.}, 1:388--429, 1998.

\bibitem{Benguria:1976in}
R.~Benguria, P.~Cordero, and C.~Teitelboim.
\newblock {Aspects of the Hamiltonian Dynamics of Interacting Gravitational
  Gauge and Higgs Fields with Applications to Spherical Symmetry}.
\newblock {\em Nucl. Phys.}, B122:61--99, 1977.

\bibitem{Rovelli:1994ge}
Carlo Rovelli and Lee Smolin.
\newblock {Discreteness of area and volume in quantum gravity}.
\newblock {\em Nucl.Phys.}, B442:593--622, 1995.

\bibitem{Ashtekar:1998ak}
Abhay Ashtekar, Alejandro Corichi, and Jose~A. Zapata.
\newblock {Quantum theory of geometry III: Noncommutativity of Riemannian
  structures}.
\newblock {\em Class. Quant. Grav.}, 15:2955--2972, 1998.

\bibitem{Freidel:2011ue}
Laurent Freidel, Marc Geiller, and Jonathan Ziprick.
\newblock {Continuous formulation of the Loop Quantum Gravity phase space}.
\newblock {\em Class. Quant. Grav.}, 30:085013, 2013.

\bibitem{Lewandowski:2005jk}
Jerzy Lewandowski, Andrzej Okolow, Hanno Sahlmann, and Thomas Thiemann.
\newblock {Uniqueness of diffeomorphism invariant states on holonomy-flux
  algebras}.
\newblock {\em Commun.Math.Phys.}, 267:703--733, 2006.

\bibitem{Schwarz:1978cn}
Albert~S. Schwarz.
\newblock {The Partition Function of Degenerate Quadratic Functional and
  Ray-Singer Invariants}.
\newblock {\em Lett. Math. Phys.}, 2:247--252, 1978.

\bibitem{Blau:1989bq}
Matthias Blau and George Thompson.
\newblock {Topological Gauge Theories of Antisymmetric Tensor Fields}.
\newblock {\em Annals Phys.}, 205:130--172, 1991.

\bibitem{Ikeda:1993fh}
Noriaki Ikeda.
\newblock {Two-dimensional gravity and nonlinear gauge theory}.
\newblock {\em Annals Phys.}, 235:435--464, 1994.

\bibitem{Schaller:1994es}
Peter Schaller and Thomas Strobl.
\newblock {Poisson structure induced (topological) field theories}.
\newblock {\em Mod. Phys. Lett.}, A9:3129--3136, 1994.

\bibitem{Cattaneo:2000iw}
Alberto~S. Cattaneo and Giovanni Felder.
\newblock {Poisson sigma models and symplectic groupoids (in Oberwolfach 1999,
  Quantization of singular symplectic quotients)}.
\newblock {\em Progress in Mathematics}, 198:61--93, 2001.

\bibitem{Gotay:1978dv}
M.~J. Gotay and J.~M. Nester.
\newblock {Presymplectic Hamilton and Lagrange Systems, Gauge Transformations
  and the Dirac Theory of Constraints}.
\newblock {\em Lect. Notes Phys.}, 94:272--279, 1979.
\newblock [,272(1978)].

\end{thebibliography}

\end{document}